\documentclass{appolb}
\usepackage{epsfig}

\newcommand{\text}[1]{\mbox{\scriptsize{#1}}}
\begin{document}
\title{Isospin mixing in nuclei around $N\simeq Z$
and the superallowed $\beta$-decay%
\thanks{Presented at the International Conference on Nuclear Structure {\it Nuclear Landscape at the Limits},
Zakopane, Poland, 2010.} 
}
\author{W. Satu{\l}a$^{1,2}$, J. Dobaczewski$^{1,3}$,  W. Nazarewicz$^{4,5,
1}$, \\ and M. Rafalski$^{1}$
\address{$^{1}$Institute of Theoretical Physics,
University of Warsaw, ul. Ho\.za 69, \\ PL-00-681 Warsaw, Poland}
\address{$^{2}$ KTH (Royal Institute of Technology),
           AlbaNova University Center, \\ 106 91 Stockholm, Sweden}
\address{$^{3}$Department of Physics,  University
of Jyv\"askyl\"a, P.O. Box 35 (YFL), \\ FI-40014 Finland}
\address{$^{4}$Department of Physics \&
  Astronomy, University of Tennessee, \\  Knoxville, Tennessee 37996, USA}
\address{$^{5}$Physics Division, Oak Ridge National Laboratory, P.O. Box
  2008, \\ Oak Ridge, Tennessee 37831, USA}
}

\maketitle
\begin{abstract}
Theoretical approaches that use one-body densities as dynamical
variables, such as Hartree-Fock or the density functional
theory (DFT), break isospin symmetry both explicitly, by virtue of
charge-dependent interactions, and spontaneously. To restore
the spontaneously broken isospin symmetry,  we implemented the isospin-projection
scheme on top of the Skyrme-DFT approach. This development allows for consistent
treatment of isospin mixing in both ground and exited nuclear
states. In this study, we apply this method to evaluate the
isospin impurities in ground states of even-even and odd-odd $N\simeq
Z$ nuclei. By including simultaneous isospin
and angular-momentum projection, we compute the isospin-breaking corrections to the $0^+\rightarrow 0^+$ superallowed $\beta$-decay.
\end{abstract}

\PACS{21.10.Hw, 
21.60.Jz, 
21.30.Fe, 
23.40.Hc 
}

\section{Introduction}

The atomic nucleus is a quantum system composed of the two types of
strongly interacting fermions, the nucleons. The charge independence of the nuclear interaction
is at the roots of  the isospin
symmetry~\cite{[Hei32a],[Wig37]}. This concept remains valid even in
the presence of Coulomb interaction, which is the major source of the
isospin  breaking. This is so because of the  smallness
of the isospin-breaking isovector and isotensor components
of the Coulomb field as compared to the isospin-conserving components of the
nuclear and Coulomb forces.

The isotopic spin quantum number, $T$,
provides  strong selection rules for nuclear reactions, decays, and transitions \cite{[Wil69]}.
In particular,  the selection rules for $\beta$-decay Fermi  and  Gamow-Teller
transitions are $\Delta T =0$ and $\Delta T =0, \pm 1$, respectively,
with the exception of $T=0 \rightarrow T=0$ transitions that are
forbidden \cite{[Wig39],[Dan65]}.
The   superallowed $0^+ \rightarrow 0^+$ Fermi transitions  bridge nuclear structure with  the electroweak standard model of particle physics,
providing the most accurate estimate for the $V_{\text{ud}}$ matrix element of the
CKM matrix \cite{[Har05c],[Tow08]}; hence,  testing the CKM unitarity.
From a nuclear structure perspective, the unitarity test depends critically on a
set of theoretically calculated isospin-breaking corrections
whose precise determination poses a challenging  problem \cite{[Mil08],[Mil09],[Tow10]}.

In this work, we calculate the isospin impurities
and isospin-breaking corrections to the superallowed
Fermi decay by using a newly
developed isospin- and angular-momentum-projected DFT
approach without pairing \cite{[Raf09c],[Sat09a],[Sat10]}.
This technique takes advantage of  the ability of the mean field (MF)
to properly describe long-range polarization effects. The MF treatment  is followed by the isospin projection
to remove the unwanted spontaneous isospin mixing within MF  \cite{[Sat09a],[Eng70],[Cau80],[Cau82]}.

This paper is organized as follows.  We begin in Sec.~\ref{model} with
a short summary of our isospin- and angular-momentum-projected
DFT approach. In Sec.~\ref{isomix}, we present
applications of the isospin-projected DFT variant of the model to
the isospin mixing in the ground states (g.s.) of even-even $N$=$Z$
nuclei. Section~\ref{fermi}  discusses preliminary results
for the isospin-breaking corrections to the superallowed beta decays calculated by considering simultaneous
isospin and angular-momentum restoration.
Finally, the conclusions are contained in Sec.~\ref{summ}.

\section{Theory}\label{model}

The isospin-projected DFT  technique \cite{[Raf09c],[Sat09a],[Sat10]}
utilizes the ability of the
self-consistent MF method to properly describe the balance between
the long-range Coulomb force and the short-range nuclear interaction,
represented in this work by the Skyrme-type energy density functional (EDF).  To
remove the spurious isospin-symmetry-breaking effects, we use the
standard one-dimensional isospin projection after variation, which
allows us to decompose the Slater determinant $|\Phi\rangle$ into
good isospin states $|T,T_z\rangle$:
\begin{equation}\label{mix} |\Phi \rangle = \sum_{T\geq
|T_z|}b_{T,T_z}|T,T_z\rangle, \quad \sum_{T\geq |T_z|} |b_{T,T_z}|^2 = 1.
\end{equation}
Here, $\hat{P}^T_{T_z T_z}$ stands for the conventional
one-dimensional isospin-projection operator:
\begin{eqnarray}\label{eqn:imk}
|TT_z\rangle & = & \frac{1}{\sqrt{N_{TT_z}}}\hat{P}^T_{T_z T_z} |\Phi\rangle
\nonumber \\ & = & \frac{2T+1}{2 \sqrt{N_{TT_z}}}\int_0^\pi d\beta_T\;
\sin\beta_T \; d^{T}_{T_z T_z}(\beta_T )\; \hat{R}(\beta_T )|\Phi\rangle ,
\end{eqnarray}
where $\beta_T$ denotes the Euler angle
associated with the rotation operator $\hat{R}(\beta_T )= e^{-i\beta_T
\hat{T}_y}$ about the $y$-axis in the isospace, $d^{T}_{T_z T_z}(\beta_T )$ is
the Wigner function~\cite{[Var88]}, and $T_z =(N-Z)/2$ is the third component
of the total isospin $T$. The  normalization factors $N_{TT_z}$, or
interchangeably the expansion coefficients $b_{T,T_z}$ that encode the
isospin content of $| \Phi \rangle$, read:
\begin{eqnarray}
\label{eqn:ovr}
N_{T T_z}  & \equiv &  |b_{T,T_z}|^2 = \langle \Phi | \hat{P}^T_{T_z T_z} | \Phi \rangle
\nonumber \\
& = & \frac{2T+1}{2}\int_0^\pi d\beta_T \sin\beta_T \; d^{T}_{T_z T_z}
(\beta_T ) \; {\mathcal N}(\beta_T),
\end{eqnarray}
where ${\mathcal N}(\beta_T)  = \langle \Phi| \hat{R}(\beta_T)| \Phi\rangle$
is the so-called overlap kernel. For technical aspects concerning the
calculation of the overlap and Hamiltonian kernels, we refer the reader to
Ref.~\cite{[Sat10]}.
The isospin-projected DFT  technique utilizes the ability of
the HF solver HFODD~\cite{[Dob09d]} to produce fully
symmetry-unrestricted Slater determinants $|\Phi\rangle$.

The isospin projection determines the set of good isospin states
(called the {\it basis\/} in the following), which in the next step is used to rediagonalize
the entire nuclear Hamiltonian, consisting of the kinetic energy, Skyrme EDF,
and the isospin-breaking Coulomb force. The rediagonalization leads to the
eigenstates:
\begin{equation}\label{mix2}
|n,T_z\rangle
= \sum_{T\geq |T_z|}a^n_{T,T_z}|T,T_z\rangle ,
\end{equation}
numbered by index $n$. The amplitudes  $a^n_{T,T_z}$ define the degree of
isospin mixing through the so-called isospin-mixing
coefficients (or isospin impurities)
 for the $n-$th eigenstate:
\begin{equation}
\label{truemix}
\alpha_C^n = 1 - |a^n_{T,T_z}|_{\text{max}}^2,
\end{equation}
where $|a^n_{T,T_z}|_{\text{max}}^2$ stands for the squared norm of the dominant amplitude in the wave function
$|n,T_z\rangle$.
It is worth stressing that the isospin projection, unlike  particle-num\-ber
or angular-momentum projections, is essentially non-singular; hence, it
can be safely used with the local EDFs. The rigorous analytical proof of this useful property can be found in Ref.~\cite{[Sat10]}.

The combined isospin and angular-momentum projection
leads to the set of states,
\begin{equation}\label{ITbasis}
|I,M,K; T,T_z\rangle =   \frac{1}{\sqrt{N_{TT_z;IMK}}}
\hat P^T_{T_z T_z} \hat P^I_{MK} |\Phi \rangle ,
\end{equation}
which form another normalized basis built on  $|\Phi \rangle$.
Here, $\hat P^T_{T_z T_z}$ and $\hat P^I_{MK}$ stand for the isospin
and angular-momentum projection operators, respectively, and
$M$ and $K$ denote the angular-mo\-men\-tum components along
the laboratory and intrinsic $z$-axes, respectively \cite{[RS80]}.
Now the problem becomes
more complicated because of the overcompleteness of the basis (\ref{ITbasis})
related to the $K$-mixing. This is overcome by performing the rediagonalization
of the Hamiltonian in the so-called {\it collective space}, spanned for
each $I$ and $T$ by the {\it natural states\/}, $|IM;TT_z\rangle^{(i)}$, as
described in Refs.~\cite{[Dob09d],[Zdu07a]}.
Such a rediagonalization  gives the solutions:
\begin{equation}   \label{KTmix}
|n; IM; T_z\rangle =  \sum_{i,T\geq |T_z|}
   a^{(n)}_{iIT} | IM; TT_z\rangle^{(i)} ,
\end{equation}
which are labeled by the index $n$ and by the conserved quantum numbers $I$, $M$, and
$T_z=(N-Z)/2$ [cf.\ Eq.~(\ref{mix2})].

\section{Isospin mixing}\label{isomix}

\begin{figure}\begin{center}
\includegraphics[angle=0,width=0.50\textwidth,clip]{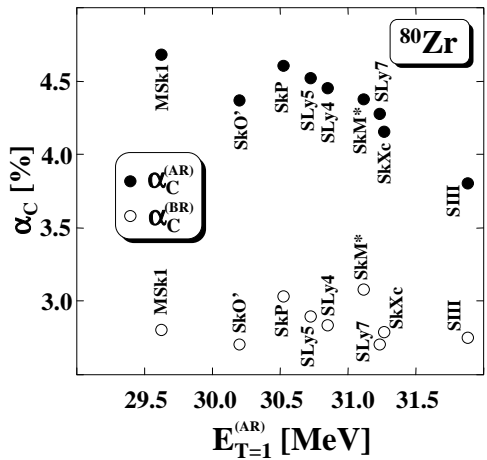}\hspace{0.05\textwidth}%
\begin{minipage}[b]{0.44\textwidth}
\caption[T]{\label{fig1}
Isospin impurities in the ground state of $^{80}$Zr, predicted by DFT, using various  Skyrme parametrizations \protect\cite{[Ben03]}
plotted versus the
corresponding excitation energies of the $T=1$ doorway states. Open dots mark
results obtained before the Coulomb rediagonalization (BR),
$\alpha_C^{(BR)} = 1 - |b_{T=|T_z|,T_z}|^2 $, which were calculated by
using expansion coefficients of Eq.~(\protect\ref{mix}). Full
dots mark the impurities (\protect\ref{truemix}) obtained after the Coulomb
rediagonalization (AR).}
\end{minipage}
\end{center}\end{figure}
By using the perturbation theory~\cite{[Sli65]} and the
analytically solvable hydrodynamical model~\cite{[Boh67]}, the
isospin mixing in atomic nuclei has been studied  since the
1960s (see Ref.~\cite{[Aue83]} for a review). These simple approaches accounted for such qualitative
features of the isospin impurities like the steady increase in $N=Z$
nuclei with increasing proton number and strong quenching with
increasing $|N-Z|$.  Quantitatively, however, their predictions for
the values of the isospin impurities $\alpha_C$ were not very
reliable.

Increased demand for accurate values of isospin mixing has been
stimulated by the recent high-precision measurements of superallowed
$\beta$-decay rates~\cite{[Har05c],[Tow08]}. Large-scale shell-model
approaches \cite{[Orm95a]}, although very accurate in the description of
configuration mixing, can hardly account for the long-range
polarization exerted on the neutron and proton states by the Coulomb
force whose accurate treatment  requires using  large configuration  spaces.
In contrast, in self-consistent DFT,
such polarization effects are naturally accounted for by finding the proper
balance between the Coulomb force, which tends to make the proton and
neutron states different, and the isoscalar part of the strong force,
which has an opposite tendency.

In general, isospin impurities determined without removing  spurious isospin
mixing are underestimated  by about 30\% compared to  the values obtained after
rediagonalization \cite{[Sat09a]}. In
the particular case of $^{80}$Zr, the removal of  spurious admixtures
increases $\alpha_C$  from $\sim$2.9\% to $\sim$4.4\%, as illustrated
in Fig.~\ref{fig1}. It is encouraging to see that the latter
value agrees well with the central value of empirical impurity
deduced from the giant dipole resonance $\gamma$-decay studies, as
communicated during this meeting by F. Camera {\it et
al.}~\cite{[Cam10]}.  Unfortunately, experimental error bars
are too large to discriminate between various Skyrme parametrizations, which
differ in predicted values of $\alpha_C$ by as much as $\sim$10\%.

\begin{figure}\begin{center}
\includegraphics[angle=0,width=1.0\columnwidth,clip]{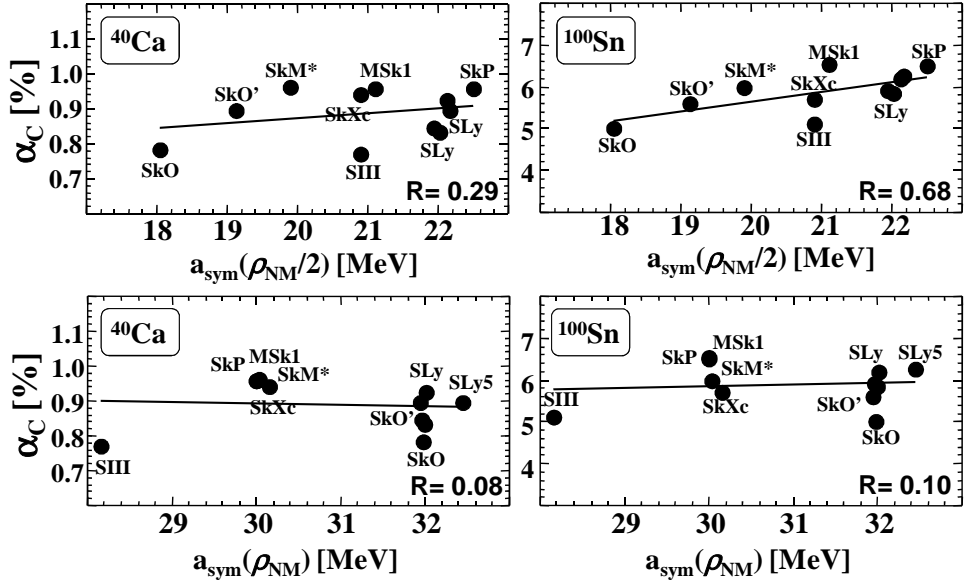}
\caption[T]{\label{fig2}
Isospin impurities predicted by  several Skyrme EDFs for the ground states of $^{40}$Ca (left) and $^{100}$Sn
(right) plotted versus
the surface (top) and volume (bottom) symmetry energy.}
\end{center}\end{figure}

Figure~\ref{fig2} illustrates our attempts to correlate the values of $\alpha_C$
with the surface and volume symmetry energies, which are primary
quantities characterizing the isovector parts of nuclear EDFs. The linear
regression coefficients shown in the figure hardly indicate any
correlation of $\alpha_C$ with these quantities. In fact, no clear correlation
was found between the calculated values of $\alpha_C$
and other bulk characteristics of the Skyrme EDFs, including
the isovector and isoscalar effective masses, and incompressibility.

\section{Isospin-breaking corrections to the Fermi
matrix elements of the superallowed $\beta$-decay}
\label{fermi}

An accurate evaluation of $\alpha_C$ is a prerequisite for
determining
the iso\-spin-breaking correction $\delta_C$ to the $0^+  \rightarrow 0^+$ Fermi
matrix element of the isospin raising/lowering operator $\hat T_{\pm}$
between nuclear states  connected by the superallowed $\beta$-decay:
\begin{equation}\label{fermime}
|\langle I^\pi=0^+, T\approx 1, T_z = \pm 1 | \hat T_{\pm}
| I^\pi=0^+, T\approx 1, T_z = 0 \rangle |^2
\equiv 2 ( 1-\delta_C ).
\end{equation}
Here, the state $| I^\pi=0^+, T\approx 1, T_z = \pm 1 \rangle$
corresponds to the g.s.\ of the even-even nucleus whereas
$|I^\pi=0^+, T\approx 1, T_z = 0 \rangle$ denotes its
isospin-analogue  in the neighboring $N=Z$ odd-odd nucleus.
Unlike the former one, the  odd-odd configuration
cannot be  expressed in a form of a MF product wave function  \cite{[Sat10]}.
Therefore, to compute the states in odd-odd $N=Z$ nuclei, we use the following strategy
(see Fig.~1 of Ref.~\cite{[Sat10]} for a schematic illustration):
\begin{itemize}
\item
Firstly, we compute the so-called antialigned g.s.\
configuration, $|\bar \nu \otimes \pi \rangle$
or  $| \nu \otimes \bar \pi \rangle$,
by placing the odd neutron and the odd proton in the lowest available
time-reversed (or signature-reversed) single-particle Nilsson orbits.
\item
Secondly, to correct for the fact that
the antialigned configurations manifestly break the isospin symmetry, that is,
$|\bar \nu \otimes \pi \rangle \approx \frac{1}{\sqrt 2} (|T=0 \rangle
+ |T=1 \rangle )$, we apply the simultaneous isospin and angular-momentum projection to create
the good isospin and good angular momentum basis $|I,M,K,T,T_z=0 \rangle$
of Eq.~(\ref{ITbasis}).
\item
Finally, to obtain the state $| I=0, T\approx 1, T_z = 0 \rangle$, we rediagonalize the total Hamiltonian, including the
Cou\-lomb term, in
the new  basis  [cf.\ Eq.~(\ref{KTmix})].
\end{itemize}
The projected $| I^\pi=0^+, T\approx 1, T_z = \pm 1 \rangle$ states in even-even nuclei are computed in the same way.

\begin{figure}\begin{center}
\begin{minipage}[b]{0.48\textwidth}
\begin{center}
\includegraphics[angle=0,width=0.95\textwidth,clip]{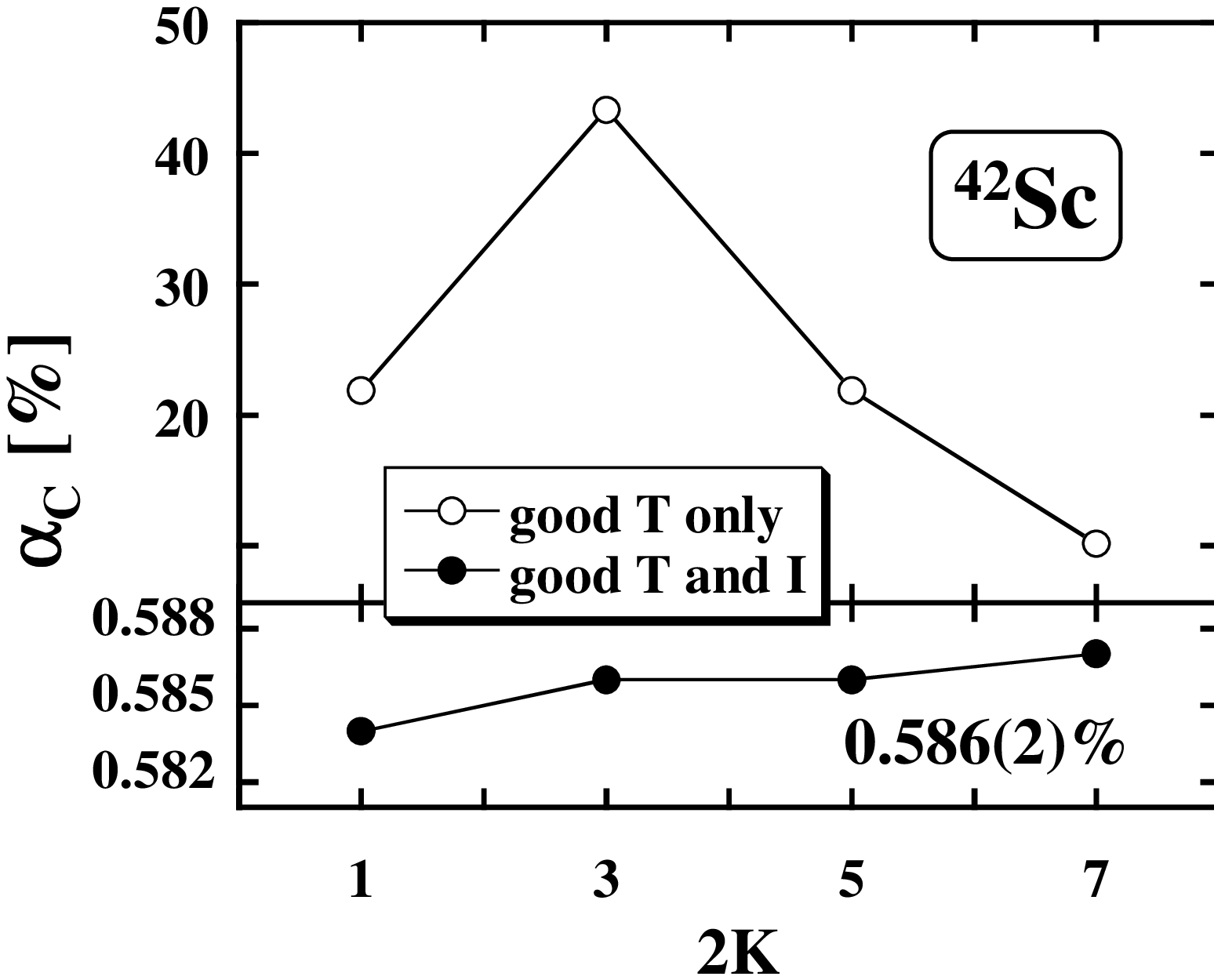}
\caption[T]{\label{fig3}
Isospin impurities in $^{42}$Sc, calculated for four antialigned
configurations that are obtained by putting the valence neutron and proton in
opposite-$K$ Nilsson orbitals originating from the $f_{7/2}$ shell, that is,
$|\nu\bar{K} \otimes \pi K \rangle $
with $K=1/2,3/2,5/2$, and 7/2. Open and full dots show the results obtained
by employing only the isospin projection and simultaneous isospin and
angular-momentum ($I=0$) projection, respectively.
}
\end{center}
\end{minipage}\hspace{0.03\textwidth}%
\begin{minipage}[b]{0.48\textwidth}
\begin{center}
\includegraphics[angle=0,width=0.95\textwidth,clip]{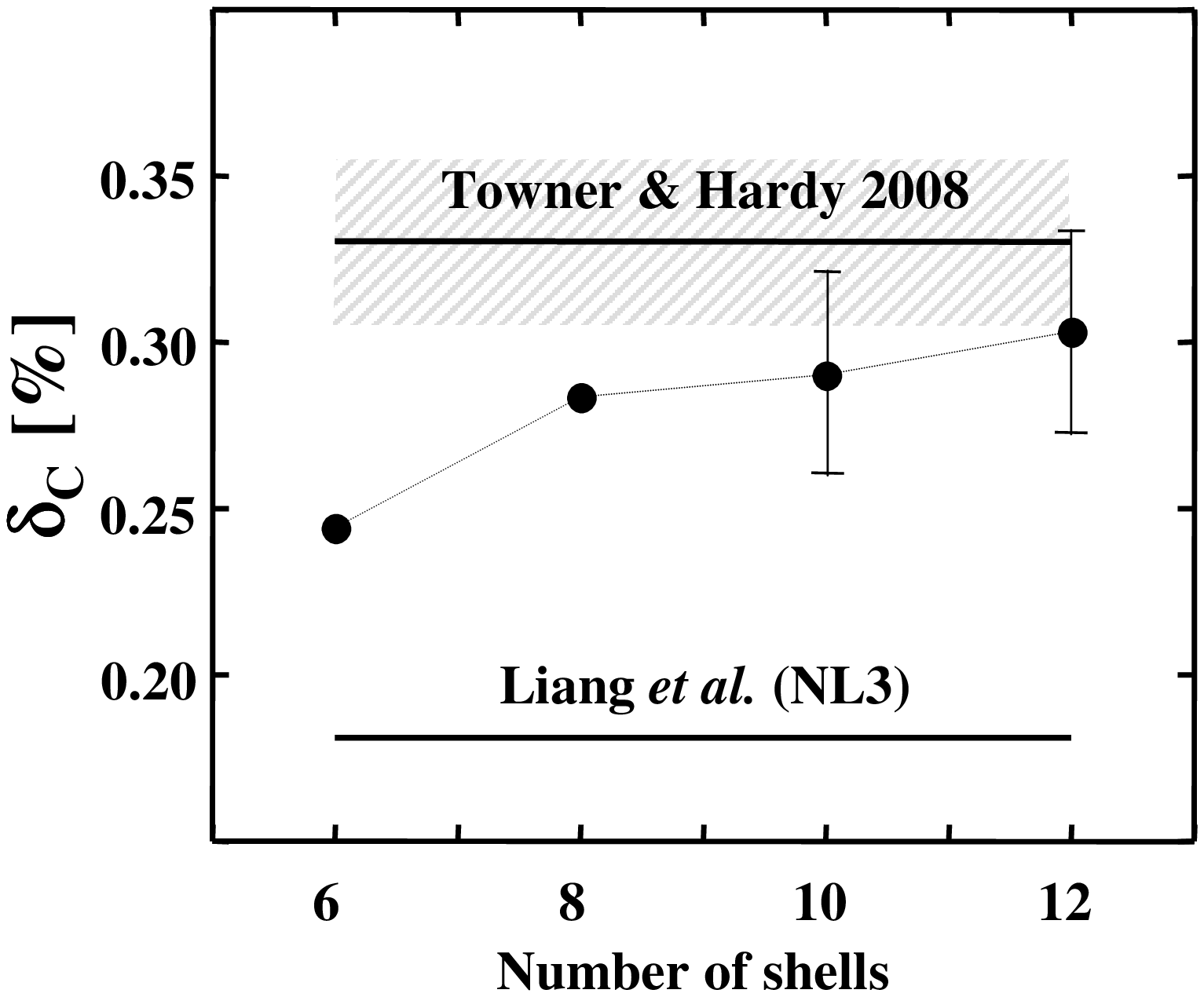}
\caption[T]{\label{fig4}
Isospin-breaking correction to the Fermi matrix element for the
superallowed transition $^{14}$O$\rightarrow$$^{14}$N. Full dots
represent our results plotted as a function of the basis size (number of  HO shells taken in HF calculations).
A conservative  $10$\%  error  was assigned to the last two points.
The values quoted in Refs.~\protect\cite{[Tow08]} (including errors)
and \protect\cite{[Lia09]} are shown for comparison. \vspace{7mm} \strut
}
\end{center}
\end{minipage}
\end{center}\end{figure}

Restoration of angular momentum turns out to be the key ingredient in
evaluation of the isospin impurity in odd-odd nuclei. This is
illustrated in Fig.~\ref{fig3}, which shows
$\alpha_C$  calculated for the $T\approx 1$ states in $^{42}$Sc. Four
solutions shown in Fig.~\ref{fig3} correspond to the four possible
antialigned MF configurations built on the Nilsson orbits
originating from the spherical $\nu f_{7/2}$ and $\pi f_{7/2}$
subshells. These  configurations can be labeled in terms of
the $K$ quantum numbers, $K=1/2$, 3/2, 5/2, and 7/2, as $|\nu\bar{K}
\otimes \pi K \rangle $.
In a simple shell-model picture, each of those MF states contains all $I$=0, 2, 4, and 6
components.
From the results shown in Fig.~\ref{fig3},
it is evident that the isospin projection alone (upper panel) leads
to unphysically large impurities, whereas the impurities obtained
after the isospin and angular-momentum ($I$=0) projection (lower panel with the scale
expanded by the factor of 500) are essentially independent of the
initial MF configuration, as expected. The average value and standard
deviation of 0.586(2)\% shown in the figure were obtained for
the configuration space of $N=10$ spherical harmonic oscillator (HO) shells,
whereas for $N=12$ the analogous result is 0.620(2)\% (see below).

Although indispensable, the angular-momentum projection creates
numerous practical difficulties when applied in the context of DFT, that is, with energy functional rather than Hamiltonian. The major problem is the presence
of singularities in energy kernels \cite{[Zdu07]}. Although appropriate
regularization schemes have already been proposed \cite{[Lac09]},
they have  neither been tested nor implemented. This fact narrows the
applicability of the model only to those EDF parametrizations which
strictly correspond to an interaction, wherefore the singularities do
not appear. For  Skyrme-type functionals, this leaves only  one EDF parametrization, namely SV \cite{[Bei75]}. This specific
EDF contains no density dependence and, after including
all tensor terms in both time-even and time-odd channels, it can be related
to a  two-body interaction.
Despite the fact that for basic observables and
characteristics such as  binding energies, level densities, and symmetry
energy, SV performs  poorly, we have
decided to use it in our systematic calculations of  $\delta_C$. Indeed,
while SV would not be our first  choice for nuclear structure predictions, it is still expected  to capture  essential polarization effects due to the self-consistent
balance between the  long-range Coulomb and short-range nuclear forces.

In order to test the
performance of our model, we have selected the superallowed $\beta$-decay transition
$^{14}$O$\longrightarrow$$^{14}$N. This case is particularly simple,
because ({\it i\/}) the participating nuclei are spherical and almost
doubly magic, which implies suppressed pairing correlations,  and ({\it
ii\/}) the antialigned configuration in $^{14}$N involves a single  $|\nu\bar{p}_{1/2} \otimes \pi p_{1/2} \rangle $ configuration that is
uniquely defined.
The predicted values of  $\delta_C$
are shown in
Fig.~\ref{fig4} as a function of the assumed configuration space
(that is, the number of spherical HO shells $N$
used). While the full convergence has not yet been achieved,  this
result, taken together with other
tests performed for heavier nuclei,  suggests that at least $N=10$
shells are needed for light nuclei ($A< 40$), whereas at least
$N=12$ shells are required for heavier nuclei. The resulting  systematic error due to basis cut-off  is estimated at the level of $\sim$10\%.

Even though calculations for all heavy ($A> 40$) nuclei
of interest are yet to be  completed, and  due to the shape-coexistence effects there are still some ambiguities concerning the choice of
global minima, our very preliminary results are
encouraging. Namely, the mean value of the structure-independent
statistical-rate function $\bar{{\cal F}}t$, obtained for 12 out of
13 transitions known empirically with high precision (excluding
$^{38}$K$\rightarrow$$^{38}$Ar case), equals $\bar{{\cal F}}t = 3069.4(10) $,
which gives the $V_{\text{ud}} = 0.97463(24) $ amplitude of the CKM matrix. These
values
 match very well those obtained by Towner and Hardy in their
latest compilation~\cite{[Tow08]}. That said,
owing to the  poor  quality of the SV parameterization, the confidence
level~\cite{[Tow10]} of our results is low. On a positive note,
 our method is quantum mechanically consistent (see
discussion in Ref.~\cite{[Mil08]}) and contains no free parameters.

\section{Summary}\label{summ}

In summary, the isospin- and angular-momentum-projected DFT theory has been
employed  to  calculate isospin mixing and isospin-breaking
corrections to the $0^+ \rightarrow 0^+$ Fermi superallowed $\beta$-decay.
Our parameter-free  model capitalizes on  the ability of the  MF approach to describe long-range polarization
effects. The  self-consistent HF wave functions containing essential correlations due to the symmetry-breaking mechanism are then used as trial states during the  projection procedure.  The results for $\alpha_C$
in $^{80}$Zr are consistent with current experimental estimates from the giant dipole resonance studies.
The preliminary results on the $\delta_C$-corrections
are also very encouraging.
The calculated values of the nucleus-independent
$\bar{{\cal F}t}=  3069.4(10)$ and the $V_{\text{ud}}= 0.97463(24) $ are consistent with
the recent evaluations of  Ref.~\cite{[Tow08]}.

This work was supported in part by the Polish Ministry of Science
under Contracts numbers N~N202~328234 and N~N202~239037, Academy of Finland and
University of Jyv\"askyl\"a within the FIDIPRO programme, and by the Office of
Nuclear Physics,  U.S. Department of Energy under Contract Nos.
DE-FG02-96ER40963 (University of Tennessee) and
DE-FC02-09ER41583  (UNEDF SciDAC Collaboration).
We acknowledge the CSC - IT Center for Science Ltd, Finland for the
allocation of computational resources.


\end{document}